\documentstyle[aps,twocolumn]{revtex}

\newlength{\extraspace}
\newlength{\extraspaces}
\catcode`\@=11
 
\def\numberbysection{\@addtoreset{equation}{section}
\def\theequation{\arabic{section}.\arabic{equation}}}

\newcommand{\be}{\begin{equation}
\addtolength{\abovedisplayskip}{\extraspaces}
\addtolength{\belowdisplayskip}{\extraspaces}
\addtolength{\abovedisplayshortskip}{\extraspace}
\addtolength{\belowdisplayshortskip}{\extraspace}}
\newcommand{\ee}{\end{equation}}
\newcommand{\ba}{\begin{eqnarray}
\addtolength{\abovedisplayskip}{\extraspaces}
\addtolength{\belowdisplayskip}{\extraspaces}
\addtolength{\abovedisplayshortskip}{\extraspace}
\addtolength{\belowdisplayshortskip}{\extraspace}}
\newcommand{\ea}{\end{eqnarray}}
\newcommand{\nonu}{\nonumber \\[.5mm]}

\begin{document}
\begin{center}
%{\huge{\bf 
{\large{\bf 
Supersymmetric Unified Models }} \\
\end{center}
\begin{center}
Lecture given at the KOSEF-JSPS Winter School, \\
Recent Developments in Particle and Nuclear Theory \\
February 21- March 2, 1996,  \\
{\sc Norisuke Sakai, Tokyo Institute of Technology}
 \\
\end{center}
%

%\LARGE
%\huge
%
\begin{enumerate}
\item {\bf Motivations for Supersymmetry}
\begin{itemize}
\item Gauge Hierarchy 
\item Higgs Scalar
%Technicolor and Supersymmetry 
%\item Unified Theories Including Gravity 
\end{itemize}
\item {\bf Introduction to Supersymmetry}
\begin{itemize}
\item Spinors %and Grassmann Number 
\item Supertransformation 
\item Unitary Representation 
%of Supersymmetry Algebra
\item Chiral Scalar Superfield 
\item Supersymmetric Field Theory 
%and Superpotential 
%\item Supersymmetric Gauge Theory 
%\item Nonrenormalization Theorem 
%\item Supersymmetry Breaking 
\end{itemize}
\item {\bf Supersymmetric $SU(3)\times SU(2)\times U(1)$ Model}
\begin{itemize}
\item Yukawa Coupling 
\item Particle Content %and R-Parity 
\end{itemize}

\end{enumerate}

%\vfill \eject

\numberbysection
\setcounter{equation}{0}

\section{ Motivations for Supersymmetry}
\subsection{Gauge Hierarchy }
\begin{enumerate}
\item {Standard model}

All the available experimental data at low energies ($E < 100 $GeV) 
can be adequately described by the standard model with $SU(3) \times 
SU(2) \times U(1)$ gauge group. 
The three different gauge coupling constants originates from the three 
different interactions, namely, strong, weak and electormagnetic 
interactions. 
The standard model has many parameters which have to be 
measured by experiments. 
There are also other conceptually unsatisfactory points as well. 
For instance, the electric charge is found to be quantized in nature, 
but this phenomena is just an accident in the standard model. 
\item {Grand unified theories}

The three interactions described by the three different gauge groups 
can be truly unified into a single gauge group if we choose a simple 
gauge group to describe all three interactions. 
This is realized by the grand unified theories proposed by Georgi and 
Glashow \cite{GG}. 
The grand unified theories achieved at least two good points:
\begin{itemize}
\item 
Because of simple gauge group, the electromagnetic charge is now 
quantized. 
\item 
Since it unifies all three couplings at high energies, 
it gives one constraint for three couplings. 
Therefore it predicts the Weinberg angle $\theta_W$. 
the prediction with a simplest possibility was found to be 
not very far from the experimental data. 
On the other hand, the unification energy $M_G$ is now very large 
compared to the electroweak mass scale $M_W$ \cite{GQW}
\be
{M_W^2 \over M_G^2} \approx \left({10^2 \over 10^{16}}\right)^2 
\approx 10^{-28}
\label{gutscale}
\ee
\end{itemize}
\item{Gravity}

Even if one do not accept the grand unified theories, one is sure 
to accept the existence of gravitational interactions. 
The mass scale of the gravitational interactions is given by the Planck 
mass $M_{Pl}$ 
\be
{M_W^2 \over M_{Pl}^2} \approx \left({10^2 \over 10^{19}}\right)^2 
\approx 10^{-34}
\label{planckscale}
\ee

Now we have a problem of how to explain these extremely small ratios 
between the mass squared $M_W^2$ to the fundamental mass squared 
$M_G^2$ or $M_{Pl}^2$ in eq.(\ref{gutscale}) or eq.(\ref{planckscale}). 
This problem is called the {\bf gauge hierarchy problem}. 
\end{enumerate}

\subsection{Higgs Scalar}

When we say {\bf explain}, we mean that it should be given a symmetry
 reason. 
This principle is called the naturalness hypothesis \cite{V}, \cite{TV}. 
More precisely, the system should acquire higher symmetry as we 
let the small parameter going to zero. 
The examples of the enhanced symmetry corresponding to the 
small mass parameter are 
\ba
m_{J=1/2} \rightarrow 0 \quad \Leftrightarrow  & 
\quad {\rm Chiral} \ \ {\rm symmetry} \nonu
m_{J=1} \rightarrow 0 \quad \Leftrightarrow  & 
\quad {\rm Local} \ \ {\rm gauge} \ \ {\rm symmetry} 
\ea

The electroweak mass scale $M_W$ originates from the vacuum 
expectation value $v$ of the Higgs field. 
The scale of $v$ in turn comes from the quadratic term of the 
higgs potential, namely the (negative) mass squared of the Higgs 
scalar $\varphi$. 
Therefore we need to give symmetry reasons for the vanishing Higgs 
scalar mass in order to explain the gauge hierarcy problem. 

Classically the vanishing mass for scalar filed 
does give rise to an enhanced symmetry called scale invariance. 
However, it is well-known that the scale invariance cannot be maintained 
quantum mechanically. 
Therefore we have only two options to explain the gauge hierarchy 
problem. 
\begin{enumerate}
\item {Technicolor model} \cite{Susskind} 

We can postulate that there is no elementary Higgs scalar at all. 
The Higgs scalar in the standard model has to be provided as a composite 
field at low energies. 
This option requires nonperturbative physics already at energies 
of the order of TeV. 
It has been rather difficult to construct realistic models which pass 
all the test at low energies especially the absence of flavor-changing 
neutral currrent. 
Models with composite Higgs scalar are called Technicolor models. 
\item {Supersymmetry} \cite{Sakai}

Another option is to postulate a symmetry between Higgs scalar and a 
spinor field. 
Then we can postulate chiral symmetry for the spinor field to make it 
massless. 
The Higgs scalar also becomes massless because of the symmetry beween 
the scalar and the spinor. 
This symmetry between scalar and spinor is called supersymmetry 
\cite{WB}, \cite{NILLES}, \cite{F}. 
Contrary to the Technicolor models, we can construct supersymmetric 
models which can be treated perturbatively up to extremely high 
energies along the siprit of the 
grand unified theories. 
\end{enumerate}

%\subsection{Unified Theories Including Gravity}

\section{ Introduction to Supersymmetry}
\subsection{Spinors %and Grassmann Number 
}
%\noindent
\subsubsection{Convention} 
Metric 
($\eta_{\mu \nu}^{\rm here}
=\eta_{\mu\nu}^{\rm Wess-Bagger}
=-\eta_{\mu \nu}^{\rm Bjorken-Drell}
$)
\be
\eta_{\mu \nu}=
\left(
\begin{array}{cccc}
-1 & 0 & 0 & 0 \\
 0 & 1 & 0 & 0 \\
 0 & 0 & 1 & 0 \\
 0 & 0 & 0 & 1
\end{array}
\right)
\label{minkowskimetric}
\ee
\be
P^{\mu}=(P^0, P^1, P^2, P^3), 
\ee
\be
P_{\mu}=(-P^0, P^1, P^2, P^3)
\ee
\ba
P\cdot Q&   =&   P^{\mu}\eta_{\mu \nu}Q^{\nu} \nonu
&   =&   
- P^0 Q^0 + P^1Q^1 +P^2Q^2 +P^3Q^3 
\ea
%
%\noindent
$\gamma$ matrix ($\gamma^{\rm here}_{\mu}
=\gamma^{\rm Wess-Bagger}_{\mu}
=\gamma^{\rm Bjorken-Drell}_{\mu}$)
\be
\gamma_{\mu} \gamma_{\nu}+\gamma_{\nu} \gamma_{\mu}=-2\eta_{\mu \nu}
\label{gammamatrix}
\ee

\noindent
Chiral $\gamma$ matrix
\ba
\gamma_5
&   
=
&   
\gamma^5=i\gamma^0\gamma^1\gamma^2\gamma^3 \nonu
&   
=
&   
 \gamma_5^{\rm Bjorken-Drell}
=i \gamma_5^{\rm Wess-Bagger}
\label{gamma5def}
\ea

Antisymmetric product of $gamma$ matrices
\be
\gamma^{\mu\nu}\equiv [\gamma^{\mu}, \gamma^{\nu}]/2
\ee
\noindent
Charge conjugation matrix
\be
C^{-1} \gamma^{\mu} C=- \gamma^{\mu T}, 
\ee
%\quad  
\be
C^T=- C, 
\ee
\be
C^\dagger C=1
\ee

\noindent
Chiral projection
\be
\psi = \psi_L + \psi_R=
\psi_- + \psi_+
\ee
\be
\psi_- = P_-\psi\equiv{1-\gamma_5 \over 2}\psi, 
\ee
\be
\psi_+ = P_+\psi\equiv{1+\gamma_5 \over 2}\psi 
\ee

\noindent
Charge conjugate spinor
\be
\psi^c\equiv C \bar \psi^T, 
\ee
\be
\bar \psi\equiv \psi\gamma_0
\ee
\be
(\psi^c)_{\mp} = 
(\psi^c_{\pm}) = 
C \overline{\psi_{\pm}}^T, 
\ee
\be
\overline{(\psi^c)_{\pm}} =
\overline{(\psi^c_{\mp})} =
- \psi_{\mp}^T C^{-1}. 
\ee

\noindent
Majorana Spinor
\be
\psi^c = \psi \rightarrow \bar \psi=-\psi^T C^{-1}
\ee
%\be
%\psi_{\mp} = C \overline{\psi_{\pm}}^T, \qquad 
%\overline{\psi_{\pm}} =- \psi_{\mp}^T C^{-1}. 
%\ee

\subsubsection{Bilinear Covariants of Majorana Spinors}
\be
\bar \psi_1\psi_2=
\bar \psi_2\psi_1, 
\ee
\be
%\quad 
\bar \psi_1\gamma^{\mu}\psi_2=
-\bar \psi_2\gamma^{\mu}\psi_1, 
\ee
\be
%\quad 
\bar \psi_1\gamma^{\mu\nu}\psi_2=
-\bar \psi_2\gamma^{\mu\nu}\psi_1, 
\ee
\be
\bar \psi_1\gamma_5\gamma^{\mu}\psi_2=
\bar \psi_2\gamma_5\gamma^{\mu}\psi_1, 
%\quad 
\ee
\be
\bar \psi_1\gamma_5\psi_2=
\bar \psi_2\gamma_5\psi_1
\ee
\be
{\rm If} \ \ \psi_1=\psi_2 \rightarrow 
\bar \psi\gamma^{\mu}\psi=
\bar \psi\gamma^{\mu\nu}\psi=0
\ee

\subsubsection{Derivative of Grassmann Number} 
\be
{\partial \over \partial \psi_{\alpha}}\psi_{\beta}
=\delta_{\alpha \beta},
\ee
\be
{\partial \over \partial \bar \psi_{\alpha}} \bar \psi_{\beta}
=\delta_{\alpha \beta}
\ee
\be
{\partial \over \partial \psi_{\alpha}} \bar \psi_{\beta}
=(C^{-1})_{\beta \alpha},
\ee
\be
%\qquad
{\partial \over \partial \bar \psi_{\alpha}} \psi_{\beta}
=(C)_{\beta \alpha}
\ee
\be
{\partial \over \partial \psi_{\alpha}} 
={\partial \over \partial \bar \psi_{\beta}} 
(C^{-1})_{\beta \alpha},
\ee
%\qquad
\be
{\partial \over \partial \bar \psi_{\alpha}} 
=-
(C)_{\alpha\beta }
{\partial \over \partial \psi_{\beta}} 
\ee
\be
\bar \epsilon {\partial \over \partial \bar \theta} 
=-{\partial \over \partial \theta} \epsilon
\ee

\subsubsection{Weyl Basis %Chiral Representation 
and Two Component %Notation 
Spinor} 
%Chiral representation 
Weyl basis of $\gamma$ matrix 
\be
\gamma_0=\left(
\begin{array}{cc}
0 & 1 \\
1 & 0 \\
\end{array}
\right), 
\ee
%\qquad
\be
\gamma_j=\left(
\begin{array}{cc}
0 & \sigma^j \\
-\sigma^j & 0 \\
\end{array}
\right), 
\ee
\be
%\qquad 
\gamma_5=\left(
\begin{array}{cc}
-1 & 0 \\
0 & 1 \\
\end{array}
\right) 
\ee
\be
C=-i\gamma_2\gamma_0=\left(
\begin{array}{cc}
-i\sigma^2 & 0 \\
0 & i\sigma^2 \\
\end{array}
\right)
=\left(
\begin{array}{cc}
\epsilon_{\alpha\beta} & 0 \\
%Kugo convention -\epsilon_{\alpha\beta} & 0 \\
0 & \epsilon^{\dot\alpha \dot\beta} \\
\end{array}
\right) 
\ee

\noindent
Two component spinor notation
\be
\psi
=\left(
\begin{array}{c}
\xi_{\alpha} \\
\eta^{*\dot\alpha} \\
\end{array}
\right), 
\ee
%\quad 
\be
\epsilon_{\alpha \beta}\epsilon^{\beta\gamma}=\delta_{\alpha}^{\gamma}
\ee
\be
\bar \psi
=
\left(
\begin{array}{cc}
(\eta^{*\dot \alpha})^* &
(\xi_{\alpha})^* \\
\end{array}
\right) 
=
\left(
\begin{array}{cc}
\eta^{\alpha} &
\xi^*{}_{\dot \alpha} \\
\end{array}
\right) 
\ee
\be
\psi_-
=\left(
\begin{array}{c}
\xi_{\alpha} \\
0 \\
\end{array}
\right) 
\ee
%\qquad
\be
\psi_+
=\left(
\begin{array}{c}
0 \\
\eta^{*\dot\alpha} \\
\end{array}
\right) 
\ee
\be
\psi^c
\equiv 
C\bar \psi^T
=\left(
\begin{array}{c}
\epsilon_{\alpha\beta}\eta^{\beta} \\
\epsilon^{ \dot \alpha \dot \beta} \xi^*_{\dot\beta} \\
\end{array}
\right) 
=\left(
\begin{array}{c}
\eta_{\alpha} \\
\xi^{*\dot\alpha} \\
\end{array}
\right) 
\ee

\noindent
Majorana spinor in the Weyl basis 
\ba
\psi
&  =&  
\left(
\begin{array}{c}
\xi_{\alpha} \\
\epsilon^{\dot \alpha \dot \beta} \xi^*{}_{\dot\beta} \\
\end{array}
\right) 
=\left(
\begin{array}{c}
\xi_{\alpha} \\
\xi^{*\dot\alpha} \\
\end{array}
\right) 
%\nonu
%&  =&  
%\left(
%\begin{array}{c}
%\epsilon_{\alpha \beta} \eta^{*\beta} \\
%\eta^{*\dot\alpha} \\
%\end{array}
%\right) 
%=\left(
%\begin{array}{c}
%\eta^*{}_{\alpha} \\
%\eta^{*\dot\alpha} \\
%\end{array}
%\right) 
\ea
\ba
\bar \psi
&  =&  
\left(
\begin{array}{cc}
\epsilon^{\alpha\beta}\xi_{\beta} & \xi^*{}_{\dot \alpha} \\
\end{array}
\right) 
=\left(
\begin{array}{cc}
\xi^{\alpha} & \xi^*{}_{\dot \alpha} \\
\end{array}
\right) 
%\nonu
%&  =&  
%\left(
%\begin{array}{cc}
%\eta^{*\alpha} & \eta^{\dot \beta}\epsilon_{\dot\beta \dot\alpha} \\
%\end{array}
%\right) 
%=\left(
%\begin{array}{cc}
%\eta^{*\alpha} & -\eta_{\dot\alpha} \\
%\end{array}
%\right) 
\ea
%\be
%(\xi_{\alpha})^* \equiv 
%\bar \xi_{\dot\alpha}, \qquad  
%(\eta^{\dot\alpha})^*  \equiv 
%\bar \eta^{\alpha}
%\ee
\be
\xi^{\alpha} \equiv 
\epsilon^{\alpha\beta} \xi_{\beta}, \qquad  
\eta_{\dot\alpha}  \equiv 
\epsilon_{\dot\alpha \dot\beta} \eta^{\dot\beta}
\ee

\subsubsection{Fierz Identity for Chiral Spinor} 
\be
\theta_{\mp\alpha} \overline{\theta_{\pm}}_{\beta}=
{-1 \over 2} \overline{\theta_{\pm}}\theta_{\mp} 
\left({1 \mp \gamma_5 \over 2}\right)_{\alpha\beta}
\ee

%
%\vfill \eject
%\vskip 1.5cm

\subsection{Supertransformation } 
\subsubsection{Superfield}
Distinction between bosons and fermions by $\theta$ 

$\rightarrow$ 
$x^{\mu}, \theta$ as coordinates in superspace 

\noindent
Superfield $=$ field in superspace $\rightarrow$ $16$ component fields 
\ba
&   &   \Phi (x, \theta)
= C(x)+\bar\theta \psi(x) 
- {1 \over 2}\bar\theta \theta N(x) 
-
{i \over 2}\bar\theta \gamma_5 \theta M(x) 
\nonu
&   
&   
-
{1 \over 2}\bar\theta \gamma^{\mu}\gamma_5 \theta v_{\mu}(x) 
%\nonu
%&   
+
%&   
i\bar\theta \theta \bar \theta \gamma_5 \lambda (x) 
+{1 \over 4}(\bar\theta \theta)^2 D (x) 
\ea

\subsubsection{Supertransformation } 
\be
\delta\theta=\epsilon, \qquad 
\delta x^{\mu}=-i\bar \epsilon\gamma^{\mu}\theta
\ee
%\vfill \eject
%
\ba
\delta\Phi(x,\theta) 
&   
=
&   
\bar\epsilon 
\left({\partial \over \partial \bar\theta}-i\gamma^{\mu}\theta 
{\partial \over \partial x^{\mu}}
\right)
\Phi(x,\theta) \nonu 
&   
=
&   
- \left({\partial \over \partial \theta}-i\bar\theta\gamma^{\mu} 
{\partial \over \partial x^{\mu}} \right) \epsilon \ \Phi(x,\theta)
\nonu
&   \equiv&   
 \left[\Phi(x,\theta), \bar\epsilon Q \right] 
= \left[\Phi(x,\theta), \bar Q \epsilon \right] 
\ea

\subsubsection{Supersymmetry Algebra}
\ba
&     &    
\left[\Phi, [\bar\epsilon_1 Q, \bar Q \epsilon_2 ] \right] 
=
\left[\Phi, [\bar\epsilon_1 Q, \bar\epsilon_2 Q] \right] 
\nonu
&   =&   
\left[ [ \Phi, \bar\epsilon_1 Q], \bar\epsilon_2 Q \right]
-\left[ [ \Phi, \bar\epsilon_2 Q], \bar\epsilon_1 Q \right] \nonu
&   =&    
\left(\delta(\epsilon_2)\right) 
\left(\delta(\epsilon_1)\right)  \Phi
-
\left(\delta(\epsilon_1)\right) 
\left(\delta(\epsilon_2)\right)  \Phi \nonu
&   =&    
\left[
\left(-{\partial \over \partial \theta}+i\bar\theta\gamma^{\mu} 
\partial_{\mu}\right)
\epsilon_2
, 
\bar
\epsilon_1 
\left({\partial \over \partial \bar\theta}-i\gamma^{\nu}
\theta \partial_{\nu}\right)
\right]
 \Phi(x,\theta) \nonu
&   =&    
2 \bar\epsilon_1 \gamma^{\mu}\epsilon_2\left(-i\partial_{\mu} 
\Phi (x, \theta) \right) \nonu
&   =&    
2 \bar\epsilon_1 \gamma^{\mu}\epsilon_2 \left[
\Phi (x, \theta), 
P_{\mu}
\right] 
\ea
Supersymmetry algebra
\be
\{ Q_{\alpha}, \bar Q_{\beta} \} = 2 (\gamma^{\mu})_{\alpha\beta} 
P_{\mu}
\ee
\be
\{ Q_{\alpha}, Q_{\beta} \} = -2 (\gamma^{\mu}C)_{\alpha\beta} P_{\mu}
\ee

\noindent
Other commutation relations
\be
[Q, P_{\mu}]=0, 
\ee
\be
%\quad \quad 
[Q_{\alpha}, J^{\mu\nu}]={i \over 2}(\gamma^{\mu\nu})_{\alpha\beta} 
Q_{\beta}
\ee
\be
[P_{\mu}, P_{\nu}]=0, 
\ee
\be
% \quad 
[P_{\mu}, J^{\nu\lambda}]=
-i(\eta^{\mu\nu}P^{\lambda}-\eta^{\mu\lambda}P^{\nu})
\ee
\ba
[J_{\mu\rho}, J^{\nu\lambda}]
&   =&   
-i(\eta^{\rho\nu}J^{\mu\lambda}+\eta^{\mu\lambda}J^{\rho\nu} \nonu
&   
-
&   
\eta^{\mu\nu}J^{\rho\lambda}-\eta^{\rho\lambda}J^{\mu\nu})
\ea

\noindent
Characteristic features of supersymmetry
\begin{enumerate}
\item 
Involving anticommutators
\item
Spacetime symmetry
\end{enumerate}
\vskip 1cm
%\vfill \eject

\subsection{Unitary Representation}
Unitary Representationof Supersymmetry Algebra 
$\rightarrow$ Physical Particle Content 

\subsubsection{Massive case}
\begin{enumerate}
\item Representation of the Poincar\'e group 

Diagonalize $P^{\mu}$ 

Standard frame $P^{\mu}=(M, 0,0,0)$

Little group $=$ Stability group of $(M,0,0,0)$ $=$ $SO(3)$

Angular momentum $j$, $z$ component $m$ 

\item Representation of $Q$ by combining $(j,m)$
\be
[Q, P_{\mu}]=0  
\ee
$P^{\mu}$  can be diagonalized 
\be
[Q_{\alpha}, J^{\mu\nu}]={i \over 2}(\gamma^{\mu\nu})_{\alpha\beta} 
Q_{\beta}
\ee
$Q$ changes $j$ and $m$ by $\pm{1 \over 2}$
\be
\{ Q_{-\alpha}, Q_{-\beta} \} = 
\{ Q_{+\alpha}, Q_{+\beta} \} = 0
\ee
\be
\{ Q_{-\alpha},  \overline{Q_{-}}_{\beta} \} =
\{ Q_{+\alpha},  \overline{Q_{+}}_{\beta} \} 
= 2 M\delta_{\alpha \beta}
%=\{Q_{\alpha}, Q^*{}_{\dot\beta} \}= 2 M\delta_{\alpha \dot \beta}
\ee
$2$ kinds of ``fermions''
\ba
 \overline{ Q_{-}}_{\alpha},
\ \ &\alpha=1,2& \ \ {\rm annihilation} \ {\rm operator} \nonu
 Q_{-\alpha},  
\ \ &\alpha=1,2& \ \ {\rm creation} \ {\rm operator} \nonumber
\ea

\noindent
Suppose $\overline{ Q_{-}}_{\alpha}|j>=0, \ \alpha=1,2$
%\be
\be
\left(
\begin{array}{ccc}
    &  Q_{-1}|j> &            \\
|j> &           &  Q_{-1}Q_{-2}|j> \\
    &  Q_{-2}|j> &            \\
\end{array}
\right) 
=
\left(
\begin{array}{ccc}
    & j-{1 \over 2}  &   \\
 j  &                & j \\
    & j+{1 \over 2}  &   \\
\end{array}
\right) 
\ee
\begin{enumerate}
\item $j=0$ case $\Rightarrow$ Chiral scalar multiplet

\begin{tabular}{|c|c|c|}                     \hline \hline
spin $j$       &   field      &  degree of freedom 
\\ \hline
$0$ & two real scalar     &       $2$      
\\ \hline
$1/2$ & a Majorana spinor          &        $2$     
\\ \hline \hline
\end{tabular}

\item $j={1 \over 2}$ case $\Rightarrow$ Vector multiplet

\begin{tabular}{|c|c|c|}                     \hline \hline
spin $j$       &   field      &  degree of freedom 
\\ \hline
$0$ & a real scalar     &       $1$      
\\ \hline
$1/2$ & 2 Majorana spinor          &        $4$     
\\ \hline
$1$ & a real vector     &       $3$      
\\ \hline \hline
\end{tabular}
\end{enumerate}

\end{enumerate}

\subsubsection{Massless case}
Standard frame $P^{\mu}=(P, 0,0,P)$

\noindent
Little group $=$ Stability group of $(P, 0,0,P)$  
\be
E_2=(J^{12}, J^{01}+iJ^{23}, J^{20}+iJ^{13})
\ee

\noindent
Representation label by 
\be
 J^2=j(j+1) \quad {\rm and}\ \ {\rm helicity}  \ J^{12}=\pm j
\ee
\be
\{ Q_{\alpha}, \bar Q_{\beta} \}
= 2 \left(%{(1+\gamma_5) \over 2}
(\gamma_0+\gamma_3)\right)_{\alpha \beta} P
=4P\left(
\begin{array}{cccc}
0 & 0   & 1 & 0    \\
0 & 0   & 0 & 0    \\
1 & 0   & 0 & 0    \\
0 & 0   & 0 & 0    \\
\end{array}
\right) 
\ee
\be
\{ Q_{-1}, \bar Q_{-1} \}= 4 P, \qquad \bar Q_{-1}=Q_{-1}^*
\ee

Others vanishing 

\noindent
Multiplet
\be
|\lambda> \rightarrow |\lambda-{1 \over 2}>
\ee
CPT invariance
\be
(\lambda,\ \  \lambda-{1 \over 2}, \ \
-\lambda+{1 \over 2}, \ \ -\lambda )
\ee

\begin{tabular}{|c|c|c|}                     \hline \hline
highest               &   helicities      &  name of              \\ 
helicity              &    of fields      &  multiplet         
\\ \hline \hline
                      &                   & chiral scalar         \\
$\lambda={1 \over 2}$ & $({1 \over 2}, 0, 0,-{1 \over 2})$ &  multiplet 
 \\ \hline
                      &                   & vector                \\
$\lambda=1$           & $(1, {1\over 2},-{1\over 2},-1)$   &  multiplet 
\\ \hline
                      &                   & graviton-             \\
$\lambda=2$           & $(2, {3 \over 2}, -{3 \over 2}, -2)$ & 
 gravitino multi. 
\\ \hline \hline
\end{tabular}

\subsection{Chiral Scalar Superfield}
\subsubsection{Irreducible Representation}
General superfield $\Phi (x,\theta )$ contains too many components 
(8 bosons + 8 fermions) compared to the minimum number of physical 
degree of freedom given by the unitary representation (2 bosons +2 
fermions)

\noindent
One should find a constraint consistent with supersymmetry to realize 
the supersymmetry in a smaller space --- Key ingredient to construct 
field theories.

$\theta_{\alpha }$ : 4-components 

$\theta_{\pm \alpha }$ : 2-components

\noindent
If $\Phi(x,\theta)$ is independent of $\theta_{+\alpha}$ or 
$\theta_{-\alpha}$, the number of components is reduced to half.
\be
{\partial \over \partial \overline{\theta_{-}}} \Phi(x,\theta) =0
\ee
But 
\be
\left\{ {\partial \over \partial \overline{\theta_{-}}}, 
Q \right\} \not=0
\ee

\noindent
Therefore this constraint is not consistent with Supersymmetry.

\noindent
Definition of Covariant derivative
\be
D_{\alpha }\Phi (x,\theta)\equiv \left( {\partial \over \partial 
\bar \theta_{\alpha}}+i\left(\gamma^\mu \partial_\mu 
\theta \right)_{\alpha} \right)\Phi \left( x,\theta \right)
\ee
\be
\bar D_{\alpha }\Phi (x,\theta)= \left( -{\partial \over \partial 
\theta_{\alpha}}
-i\left(\bar \theta \gamma^\mu \partial_\mu \right)_{\alpha} 
\right)\Phi \left( x,\theta \right)
\ee
\ba
\{D_{\alpha },\bar Q_{\beta}\}
&   
=
&   
\left\{ {\partial \over \partial 
\bar \theta_{\alpha}}+i\left(\gamma^\mu \right)_{\alpha 
\gamma} \theta_\gamma \partial_\mu,  %\left(
 {\partial \over \partial \theta_{\beta}}-i \bar \theta_{\delta} 
(\gamma^\mu )_{\delta \beta} \partial_\mu %\right)
\right\}
\nonu
&    =&    i(\gamma^\mu)_{\alpha \gamma} \partial_\mu 
\left\{ \theta_\gamma ,{ \partial \over \partial \theta_\beta} \right\}
- i\left\{ {\partial \over \partial \bar \theta_\alpha},
\bar \theta_\delta \right\}(\gamma^\mu)_{\delta\beta} \partial_\mu
\nonu
&    =&    i(\gamma^\mu)_{\alpha \gamma} \partial_\mu
-i(\gamma^\mu)_{\alpha \gamma} \partial_\mu=0 
%\\
\ea
%&  &  %\Leftrightarrow 
\be
\{D_\alpha ,Q_\beta\}=0 \qquad \qquad 
\ee
\ba
\{D_\alpha , \bar D_\beta \}&   =&   
\left\{ {\partial \over \partial 
\bar \theta_\alpha }+i(\gamma^\mu)_{\alpha \gamma}
\theta_\gamma 
\partial_\mu, -{\partial \over \partial \theta_\beta}
-i\bar \theta_\delta (\gamma^\mu)_{\delta \beta}\partial_\mu 
\right\}
\nonu
&   =&   -2i(\gamma^\mu)_{\alpha \beta} \partial_\mu 
%\equiv -2(\gamma^\mu)_{\alpha \beta} P_\mu
\ea
$D_{\alpha}$ satisfies the same algebra as $Q_{\alpha}$

\subsubsection{Chiral Projected Covariant Derivative}
\be
D_{\pm \alpha}={\partial \over \partial \overline{\theta_{\mp}}_\alpha}
+i\left( \gamma^\mu \partial_\mu \theta_\mp \right)_\alpha
\ee
\be
\{D_{+\alpha},D_{+\beta}\}=\{D_{-\alpha},D_{-\beta}\}=0
\ee
\be
\{D_{\pm \alpha},D_{\mp \beta}\}
=\left({1 \pm \gamma_5 \over 2}\gamma^\mu 
C\right)_{\alpha \beta} (-2i\partial_\mu )
\ee

\noindent
Negative chiral scalar superfield
\be
D_{+\alpha} \Phi (x,\theta)=0
\ee

\noindent
Define
\ba
z^\mu &  \equiv&   x^\mu +{i \over 2}\overline{\theta}\gamma^\mu 
\gamma_5 \theta \nonu
&  =&   x^\mu-i\overline{\theta_-}\gamma^\mu \theta_{-}
%\nonu
%&  
=
%&  
 x^\mu+i\overline{\theta_+}\gamma^\mu \theta_+ 
\ea
 Then
\ba
 D_{+\alpha} z^\mu&  =&  \left({\partial \over \partial 
\overline{\theta_{-}}_\alpha }
+i(\gamma^\mu \partial_\mu \theta_-)_\alpha
\right)(x^\mu - i\overline{\theta_-}\gamma^\mu\theta_-)
\nonu
&  =&   i(\gamma^\mu \theta_-)_\alpha 
-i(\gamma^\mu \theta_-)_\alpha=0
\ea
\be
D_{+\alpha} \theta_{-\beta}=0,
\ee
\be
D_{+\alpha} \theta_{+\beta}
=-\left({1+\gamma_5 \over 2}C\right)_{\alpha \beta}\not=0
\ee
Changing variables
 $(x,\theta_+,\theta_-) \rightarrow (z,\theta_+, \theta_{-})$
%\ba
\be
D_{+\alpha} \Phi(x,\theta)
%&  
=
%&   
0 
%\nonu
\Longrightarrow 
\Phi 
%&   
=
%&   
\Phi (z,\theta_-)
\ee
%\ea
Namely, $\Phi$ is independent of $\theta_{+}$ if $z$ is fixed. 

\noindent
Let us denote negative chiral scalar field as $\Phi_-(z,\theta_-)$

\noindent
Negative chiral scalar field can be used as a representation space 
of supersymmetry ($\{Q,D\}=0$)

\subsubsection{Properties of chiral scalar superfield}
\ba
&   
&   
  \Phi_-(z,\theta_-)
=
{\rm e}^{{i \over 2}\overline{\theta}\gamma^\mu 
\partial_\mu \gamma_5 \theta} \Phi_-(x,\theta_-) \\
&   
=
&   
\left(A_-(z)+\sqrt{2}\overline{\theta_{+}}
\psi_-(z) + \overline{\theta_{+}} \theta_{-} F_-(z)\right) 
\nonu
&  =&  e^{{i \over 2}\overline{\theta}\gamma^\mu \partial_\mu 
\gamma_5 \theta}\left(A_-(x)+\sqrt{2}\overline{\theta_{+}}
\psi_-(x) + \overline{\theta_{+}} \theta_{-} F_-(x)\right)
\nonumber
\ea

\noindent
Chiral scalar field is complex 

\noindent
Degree of freedom of component fields 

\begin{tabular}{|c|c|c|c|}                     \hline \hline
  &   real or complex &  off-shell      &  on-shell    
\\ 
fields     & spin    &  real d.o.f. &  real d.o.f.   
\\ \hline \hline
 & complex & & \\
$A_-(x)$ & scalar & 2   & 2         \\ \hline
 &  complex & & \\
$\psi_-(x)$ & 2-comp. spinor & 4 & 2 \\ \hline
 & complex & & \\
$F_-(x)$ & aux. scalar & 2 & 0 
\\ \hline \hline
\end{tabular}

$\psi$ obeys the Dirac equation. 
On-shell d.o.f. is counted by $($Dirac$)^2=$Klein-Gordon. 

\noindent
Product of chiral scalar superfields $\Phi_{-}^1$ and $\Phi_{-}^2$
\ba
&  &  \Phi_{-}^1(z,\theta_-)\Phi_{-}^2(z,\theta_-)
\nonu
&  =&  
(A^1_-(z)+\sqrt{2}\overline{\theta_{+}}\psi^1_-(z)
+\overline{\theta_{+}}\theta_-F^1_-(z))
\nonu
&  
\times
&  
(A^2_-(z)
+\sqrt{2}\overline{\theta_{+}}\psi^2_-(z)
+\overline{\theta_{+}}\theta_-F^2_-(z))
\nonu
&  =&  A^1_-A^2_- + \sqrt{2}\overline{\theta_+}
(A^1_-\psi^2_- +\psi^1_-A^2_-) 
\nonu
&  
+
&  
\overline{\theta_{+}}\theta_-(F^1_-A^2_- 
+A^1_-F^2_-)
+2\overline{\theta_+} \psi^1_- \overline{\theta_{+}}\psi^2_-
\nonu
&  
=
&  
A^1_-A^2_- + \sqrt{2}\overline{\theta_+}(A^1_-\psi^2_- 
+\psi^1_-A^1_-)
\nonu
&  
+
&  
\overline{\theta_{+}}\theta_-(F^1_-A^2_- +A^1_-
F^2_- 
-\overline{(\psi_-^1)^c}
%+\psi^{1T}_-C^{-1}
\psi^2_-)
\ea
\ba
(\overline{\theta_+}\psi^1_-)(\overline{\theta_{+}}\psi^2_-)
&   
=
&   
(\overline{(\psi_-^1)^c}
\theta_-)(\overline{\theta_{+}}\psi^2_-)
\nonu
&   
=
&   
%(\overline{(\psi_-^1)^c})_\alpha 
%{-1 \over 2}(\overline{\theta_{+}}\theta_-)
%({1-\gamma_5 \over 2})_{\alpha \beta}\psi^2_{-\beta}
%\nonu
%&  =&  
{1 \over 2}(\overline{\theta_{+}}\theta_-)(\psi^1_-C^{-1}\psi^2_-)
%\nonumber
\ea

\noindent
Supertransformation for an ``infinitesimal'' $\epsilon$
\ba
\delta z^\mu 
&   =&   \delta x^\mu 
+ {i \over 2} \delta (\bar \theta \gamma^\mu \gamma_5\theta)
\nonu
&   
=
&   
-i\overline{\epsilon}\gamma^\mu \theta 
+i\overline{\epsilon}\gamma^\mu \gamma_5\theta
\nonu
&   
=
&   
-2i\overline{\epsilon_-}\gamma^\mu\theta_-
\ea
\ba
&   &   \delta \Phi_-(z,\theta_-)
=\left(\delta z^\mu{\partial \over \partial z^\mu}
+\delta \overline{\theta_{+}}
{\partial \over \partial \overline{\theta_{+}}}
\right)\Phi_-(z,\theta_-)\nonu
&   =&   \left(-2i\overline{\epsilon_-}\gamma^\mu \theta_- 
{\partial \over \partial z^\mu}+\overline{\epsilon_+}
{\partial \over \partial \overline{\theta_{+}}} \right)
\nonu
&   
\times
&   
(A_-(z)+\sqrt{2}\overline{\theta_{+}}\psi_-(z)
+\overline{\theta_{+}}\theta_-F_-(z))\nonu
&   =&   
\sqrt{2}\overline{\epsilon_{+}}\psi_- +2\overline{\epsilon_{+}}
\theta_-F_- -2i\overline{\epsilon_-}\gamma^\mu \theta_- \partial_\mu 
A_- 
\nonu
&   &   
-2\sqrt{2}i\overline{\epsilon_-} \gamma^\mu \theta_- 
\overline{\theta_{+}}\partial_\mu \psi_-\nonu
&   =&   
\sqrt{2}\overline{\epsilon_{+}}\psi_- +\sqrt{2}\overline{\theta_{+}}
\sqrt{2}(\epsilon_- F_- 
+i\gamma^\mu \epsilon_{-}\partial_\mu A_- )
\nonu
&   &   
+\overline{\theta_{+}}\theta_-
\sqrt{2}\overline{\epsilon_-} i\gamma^\mu \partial_\mu \psi_-
\ea

\noindent
Therefore
\ba
\delta A_- 
&    = &   \sqrt{2}\overline{\epsilon_{+}}\psi_- 
\nonu
\delta\psi_-
&    = &   \sqrt{2}(\epsilon_- F_- 
+i\gamma^\mu \epsilon_{-}\partial_\mu A_- )
\nonu
\delta F_-
&    = &   i\sqrt{2}\overline{\epsilon_-}\gamma^\mu \partial_\mu 
\psi_- 
\ea

\subsubsection{Positive Chiral Scalar Field}
\be
D_- \Phi_{+}=0,
\ee
\be
z^{*\mu}=x^\mu-{i \over 2}\overline{\theta}\gamma^\mu 
\gamma_5 \theta
\ee
\ba
\Phi_{+}&   =&   \Phi_{+}(z^*,\theta_+)\nonu
&   =&   A_{+}(z^*)
+\sqrt{2}\overline{\theta_-}\psi_{+}(z^*)+\overline{\theta_-}\theta_+
F_{+}(z^*)
\ea

\noindent
Product of positive chiral and negative chiral scalar fields is a 
general superfield (without a definite chirality)

\noindent
Complex conjugation changes the chirality
\ba
&   
&   
(\Phi_-(z,\theta_-))^* 
=
{\rm e}^{-{i \over 2}\overline{\theta}\gamma^\mu 
\partial_\mu \gamma_5 \theta} \nonu
&   
&   
\times 
(A^*_-(x)
+\sqrt{2}\overline{\theta_-}
(\psi_-)^c(x)
%C \overline{\psi_-}^T(x)
+\overline{\theta_{+}}
\theta_-F^*_-(x)) 
\ea

\subsection{Supersymmetric %Lagrangian 
Field Theory}
\subsubsection{Lagrangian with Chiral Scalar Fields}
Lagrangian invariant under supersymmetry transformation up to a total 
divergence:
\begin{enumerate}
\item Two possibilities

\begin{enumerate}
\item 
 $D$-term of general superfield $\Phi$
\be
[\Phi]_D={1 \over 8}(\overline{D}D)^2\Phi
\ee

\item 
$F_{\pm}$-term of chiral scalar superfield $\Phi_{\pm}$
\be
[\Phi_\pm]_F =-{1 \over 4}\overline{D}D \Phi_\pm
\ee
\end{enumerate}

\item Dimensional analysis
\be
[\Phi_\pm]=M^1
\ee
\be
[\theta_\alpha]=L^{{1 \over 2}}=M^{-{1 \over 2}},
\ee
\be
[D_\alpha]
=[\overline{D}_\alpha]=M^{{1 \over 2}}
\ee

\item Renormalizable Lagrangian (in 4-dimension)

operators with dimension $\le 4$.
\begin{enumerate}
\item D-type:
\be
(\overline{D}D)^2\cdot \Phi_{1+}\Phi_{2-}  
%{\rm the only possibility}
\ee
Dimension $[\overline{D}D]=M^2$

\item F-type:
\be
(\overline{D}D)
(a\Phi_- +b\Phi_{1-}\Phi_{2-}+c\Phi_{1-}\Phi_{2-}
\Phi_{3-})
\ee
Since $\overline{D}D$ has dimension $M^1$, up to third order 
polynomials of 
 chiral scalar superfields of \underline{one} chirality 
are renormalizable. 
\end{enumerate}

\item General Lagrangian with a single chiral scalar field
\be 
L=L_{\rm kin}+L_{\rm int.}
\ee
\ba
L_{kin}&=&{1 \over 32}(\overline{D}D)^2\Phi^*_- \Phi_- \nonu
&  =&  {1 \over 4}\partial^2A^*_-A_- 
-{1 \over 2}\partial_\nu A_-^*\partial^\nu A_-
 +{1 \over 4}A^*_-\partial^2 A_-
\nonu
&   + &   F^*_-F_- 
+{1 \over 2}\overline{\psi_-} i\gamma^\mu\partial_\mu\psi_- 
-{1 \over 2}\partial_\mu \overline{\psi_-} i\gamma^\mu\psi_-\nonu
&  =&  -\partial_\nu A^*_- \partial^\nu A_- 
+\overline{\psi_-} i\gamma^\mu \partial_\mu \psi_- +F^*_- F_- 
\nonu
& & +{\rm total} \ \ {\rm derivatives}
\ea
\ba
L_{\rm int.}
&   =&   -{1 \over 4}\overline{D}D\left({\sqrt{2} \over 3}
f\Phi^3_- +{m \over 2}\Phi^2_- +{\rm h.c.}\right)\nonu
&   =&   
\sqrt{2}f\left(F_-A^2_- 
%+\psi^T_- C^{-1}
-\overline{(\psi_-)^c}
\psi_- 
A_-\right)
\nonu
&   
&   
+m\left(F_-A_- 
%+{1 \over 2}\psi^T_- C^{-1}
-{1 \over 2}\overline{(\psi_-)^c}
\psi_-
\right)
%\nonu &    &   
+{\rm h.c.}
\ea

\noindent
Elimination of auxiliary fields $F$ from $L$ 

\noindent
Euler eq. for $F_-$
\be
F^*_- +\sqrt{2}fA^2_- +mA_- =0
\ee
\ba
L 
&   
\rightarrow
&   
-\partial_\nu A^*_- \partial^\nu A_- 
+{1 \over 2}\bar \psi i \gamma^\mu \partial_\mu \psi -{m \over 2}
\overline{\psi}\psi
\nonu
&   
&   
%+\left(\sqrt{2}f\psi^T_- C^{-1} 
-\left(\sqrt{2}f\overline{(\psi_-)^c}
\psi_- A_- +{\rm h.c.}\right)
\nonu
&   
&   
-|\sqrt{2}fA^2_- +mA_-|^2
\ea

%If $S=0$

$m$ : mass of a Majorana spinor $\psi$  and a complex scalar $A$

$f$ : 
 Yukawa coupling and 
$|A^2_-|^2$ coupling

\item
Feynman diagram calculation is facilitated by superfield perturbation
\be
-{1 \over 4}\overline{D}D \approx {1 \over 2}d\theta_1d\theta_2
\equiv d^2\theta
\ee
\be
{1 \over 32}(\overline{D}D)^4
\approx
{1 \over 4}d\theta_1d\theta_2d\theta_3d\theta_4
\equiv d^4\theta
\ee
%\vfill \eject
%\vskip 4cm

\end{enumerate}

\subsubsection{Supersymmetric Gauge Theory}
\begin{enumerate}
\item Gauge Transformation 

Ordinary local gauge transformation
\be
\psi(x)\rightarrow e^{-i\Lambda^a (x)T^a}\psi(x)
\ee
Supersymmetric extension
\be
\begin{array}{ccc}
{\rm matter} &       & {\rm chiral} \ {\rm scalar} \ {\rm superfield} \\
\psi(x)  & \rightarrow & \Phi_{-}(x,\theta) \\
\end{array}
\ee

$x$-dependent 
gauge function $\Lambda(x)$ is generalized to a chiral 
scalar superfield $\Lambda_{-}(x,\theta)$
\be
\Lambda(x)  \rightarrow \Lambda_{-}(x,\theta) 
\ee

Supersymmetrized local gauge transformation
\be
\Phi_{-}(x,\theta)
\rightarrow {\rm exp}(-i\Lambda^a_{-}(x,\theta)T^a)\Phi_{-}(x,\theta)
\ee
using gauge function superfield with the same chirality
\be
\Phi_{-}^\dagger (x,\theta)
\rightarrow 
\Phi_{-}^\dagger (x,\theta)
{\rm exp}(i\Lambda^a_{-}(x,\theta)^*T^a)
%\Phi_{+}(x,\theta)
%\rightarrow {\rm exp}(-i\Lambda^a_{-}(x,\theta)^*T^a)\Phi_+(x,\theta)
\ee

\item Gauge Invariant Kinetic Term for Matter Fields
\begin{enumerate}
\item A General Superfield for Gauge Boson and Gaugino
\be
e^{2g V^a T^a} 
\ee

\noindent
Gauge transformation
\be
e^{2gV^a T^a}\rightarrow 
e^{-i\Lambda^a_- (x,\theta)^* T^a}e^{2gV^a T^a}e^{i\Lambda^a_- T^a}
\ee

\item Kinetic Term for a Chiral Scalar Field $\Phi_-$
\be
L_{\rm kin.}
={1 \over 32}(\overline{D}D)^2(\Phi^\dagger_- e^{2gV^a T^a}\Phi_-)
\ee
is gauge invariant

\noindent
The general superfield $V^a$ is
dimensionless and 
real 
\be 
V^{a*}=V^a
\ee
\end{enumerate}
\end{enumerate}

\subsubsection{Gauge Transformation}
\begin{enumerate}
\item Gauge transformation in components

\noindent
U(1) case
\be
V \rightarrow V + {i \over 2g}(\Lambda_{-}-\Lambda^*_{-})
\ee

\noindent
In terms of components
\ba
V(x,\theta)
&   
\equiv 
&   
C(x)+i\overline{\theta}_{+}\chi_{-}(x)
-i\overline{\theta}_{-}\chi_{+}(x)\nonu
&   
 +
&   
 {i \over 2}\overline{\theta}_{+}\theta_{-}(M+iN)
-{i \over 2}\overline{\theta}_{-}\theta_{+}(M-iN)
\nonu
&   
-
&   
\overline{\theta}_{+}\gamma^\mu \theta_{+}v_\mu(x)
\nonu
&
+
&
i\overline{\theta}_{+}\theta_{-}\overline{\theta}_{-}
(\lambda_+ +{i \over 2}\gamma^\mu \partial_\mu \chi_{-}) 
\nonu
&   
-
&   
i\overline{\theta}_{-}\theta_{+} \overline{\theta}_{+}(\lambda_{-} 
+{i \over 2}\gamma^\mu\partial_\mu \chi_+)
\nonu
&  +&  
{1 \over 2}\overline{\theta}_{+}\theta_{-}\overline{\theta}_{-}
\theta_{+}\left(D+{1 \over 2}\partial^2 C \right)
\ea
\ba
C 
&  
\rightarrow 
&  
C+{i \over 2g}(A_{-} -A^*_{-}) 
\nonu
\chi_{-} 
&  
\rightarrow 
&  
\chi_{-} + \sqrt{2}{1 \over 2g}\psi_{-}, 
\quad \quad 
\nonu
\chi_{+} 
&  
\rightarrow 
&  
\chi_{+} + \sqrt{2}{1 \over 2g}
%C\bar \psi^T_{-}
(\psi_{-})^c
\nonu
M 
&  
\rightarrow 
&  
M + {1 \over 2g}(F_{-}+F^*_{-}), 
\qquad 
\nonu
N 
&  
\rightarrow 
&  
N + {i \over 2g}(F_{-}-F^*_{-})
\ea
$C, \chi, M, N$ can be gauged away
\be
v^\mu 
\rightarrow 
v^\mu + {1 \over 2g}\partial^\mu(A_{-}
+A^*_{-})
\ee
$v^\mu$ is an ordinary gauge field
\be
\lambda 
\rightarrow 
\lambda
\qquad \qquad 
D 
\rightarrow 
D
\ee
$\lambda, D$ are gauge invariant. 

\item Wess-Zumino gauge

\noindent
Eliminate $C$, $\chi$, $M$, $N$ by choosing 
%a chiral scalar field 
$\Lambda_{-}$
\ba
&  
&  
V_{WZ}
=
-\overline{\theta}_+ \gamma^\mu \theta_{+}v_\mu(x)
+i\overline{\theta}_{+} \theta_{-} \overline{\theta}_{-} \lambda_{+}
\nonu
&  
&  
-
i\overline{\theta}_{-}\theta_{+}\overline{\theta}_{+}\lambda_{-}
+{1 \over 2}\overline{\theta}_{+}\theta_{-}\overline{\theta}_{-}
\theta_{+} D(x)
\\
&  
=
&  
-{1 \over 2}\overline{\theta}\gamma^\mu \gamma_5 \theta 
v_\mu(x)+i \overline{\theta}\theta\overline{\theta}\gamma_5\lambda 
%\nonu
%&  
%&  
+
{1 \over 4}(\overline{\theta}\theta)^2D(x)
\nonumber
\ea

\noindent
Wess-Zumino gauge is not manifestly supersymmetric. 

\noindent
However, particle content is most easily seen.
\end{enumerate}

\subsubsection{Supersymmetric Gauge Field Strength}
Gauge field strength = gauge covariant building block 

\noindent
$\lambda^a (x)$ is the gauge coveriant field with lowest dimension

\noindent
Derivative $\rightarrow$ $D_{\pm\alpha}$
\be
W_{--\alpha} \equiv {-1 \over 8g}(\overline{D}_- 
D_{+})\left({\rm e}^{-2gV^a T^a} D_{-\alpha} 
{\rm e}^{2gV^a T^a}\right)
\ee
The first $-$ suffix denotes negative chiral projection for the index 
$\alpha$. 

\noindent
The second $-$ suffix denotes negative chiral superfield. 
\be
D_{+}W_{--\alpha}=0
\ee

\noindent
$W_{--\alpha }$ is gauge covariant
\be
W_{--\alpha} \rightarrow {\rm e}^{-i \Lambda^{a}_{-} T^a} 
W_{--\alpha }{\rm e}^{i\Lambda^{a}_{-} T^a}
\ee

\noindent
Similarly positive chiral field strength is given by 
\be
W_{++\alpha} = {-1 \over 8g}(\overline{D_{+}}D_{-})
({\rm e}^{2gV^a T^a}D_{+\alpha}{\rm e}^{-2gV^a T^a})
\ee

\noindent
Kinetic term for vector superfield is given by 
\be
L_{\rm gauge}={-1 \over 16}\overline{D_{+}}D_{-}\left(
\overline{W_{++}}^aW^a_{--}\right)+{\rm h.c.}
\ee

\noindent
In the Wess-Zumino gauge
\ba
%{1 \over g}
&   
&   
W_{--}
=
{\rm e}^{{i \over 2}\overline{\theta}\gamma^\mu 
\partial_\mu \gamma_5 \theta}
\\
&   
&   
\times 
\biggl[ i \lambda_- -\left(D+{i \over 2}
\gamma^{\mu\nu} v_{\mu \nu}\right)\theta_{-} 
+
\overline{\theta_{+}}\theta_{-}
(\gamma^\mu \nabla_\mu \lambda_{-})\biggr]
\nonumber
\ea
\be
v_{\mu\nu}=\partial_\mu v_\nu -\partial_{\nu} v_\mu+ig[v_\mu,v_\nu]
\ee
\be
\nabla_\mu \lambda_{-}=\partial_\mu \lambda_{-}+ig[v_\mu,\lambda_{-}]
\ee
\ba
L_{\rm gauge}
&   
=
&   
{1 \over 2}\overline{\lambda_{+}}^a
(i\gamma^\mu \nabla_\mu \lambda_{+})^a
-{1 \over 8}v_{\mu\nu}^a v^{a\mu\nu}
+
{1 \over 4}D^a D^a 
\nonu
&   
&   
+{i \over 16}\epsilon^{\mu\nu\rho\tau}
v_{\mu\nu}^a v_{\rho\tau}^a 
+ {\rm h.c.}
\\
&  =&  
{1 \over 2}\overline{\lambda}^a(i\gamma^\mu \nabla_\mu \lambda)^a 
-{1 \over 4}v_{\mu\nu}^av^{a\mu\nu}
+{1 \over 2}D^a D^a
\nonumber
\ea

\noindent
$D^a$ is an auxiliary field

\subsubsection{Gauge Interaction in the Wess-Zumino Gauge}
\ba
&  
&  
L_{{\rm kin.of} \Phi_{-}}=
-(\nabla_\mu A_{-})^\dagger \nabla^\mu 
A_{-} +\overline{\psi_{-}} i \gamma^\mu \nabla_\mu \psi_{-} 
\nonu
&   &  
+F^\dagger_- F_{-}
+i\sqrt{2}g(A^\dagger_{-} T^a \overline{\psi_{+} }
\lambda_{-}^a 
-\overline{\lambda_{-}^a} \psi_+ T^a A_{-})
\nonu
&   &  
+gA^\dagger_{-} 
D^a T^a A_{-}
\ea
\be
\nabla_\mu A_{-}=\partial_\mu A_{-}+igv_\mu^a T^a A_-
\ee

\noindent
Eliminating D by Euler eq. from $L_{gauge}+L_{kin}$
\be
D^a+gA^\dagger_- T^a A_{-}=0
\ee
\ba
{1 \over 2}D^a D^a 
&  
+
&  
gA^\dagger_{-} D^a T^a A_{-}
\nonu
&  
=
&  
-{1 \over 2}
\sum_{a}
g^2|A^\dagger_{-}T^a A_{-}|^2
\ea

\noindent
This is the D-term of the scalar potential

\noindent
U(1) $\xi$ -term (Fayet-Iliopoulos term)
\be
L_{\xi}={1 \over 16}(\overline{D}D)^2 \xi V =\xi D
\ee
\be
[\xi]=M^2
\ee

\section{ Supersymmetric $SU(3)\times SU(2)\times U(1)$ Model}
\subsection{Yukawa Coupling} 
\subsubsection{Nonsupersymmetric Standard Model} 
In the nonsupersymmetric $SU(2)\times U(1)$ model, 
we have left-handed quark doublet $q_j$, the right-handed 
$u$-type quark $u_{Ri}$ and $d$-type quark $d_{Ri}$, 
left-handed lepton doublet $l_j$, the right-handed 
electron $e_{Ri}$, together with Higgs doublets. 
We shall denote the generation index by lower suffixes $i, j, \cdots $. 
We also denote the Higgs doublets to give the masses to the $u$-type 
($d$-type) quark as $\varphi_u$ ($\varphi_d$). 
We can write down the 
Yukawa interaction between quarks, leptons and Higgs fields 
in terms of the Yukawa couplings $f$ as 
\ba
&&L_{Yukawa}
\\
&
= 
&
f^{ij}_u \overline{u_{Ri}} \varphi_u^T  \varepsilon  q_j + 
f^{ij}_d \overline{d_{Ri}} \varphi_d^T  \varepsilon  q_j + 
f^{ij}_e \overline{e_{Ri}} \varphi_d^T  \varepsilon  l_j  
\nonumber
\ea
where 
\be
q_i = 
\left(
\begin{array}{c}
u_i \\
d_i \\
\end{array}
\right), 
\qquad 
l_i = 
\left(
\begin{array}{c}
\nu_i \\
e_i \\
\end{array}
\right), 
\ee
\be
\varphi_u = 
\left(
\begin{array}{c}
\varphi_u^+ \\
\varphi_u^0 \\
\end{array}
\right), 
\qquad 
\varphi_d = 
\left(
\begin{array}{c}
\varphi_d^0 \\
\varphi_d^- \\
\end{array}
\right) 
\ee
\be
%\qquad 
\varepsilon = 
\left(
\begin{array}{cc}
0 & 1 \\
-1 & 0 \\
\end{array}
\right) 
\ee
In the nonsupersymmetric model, we can choose the Higgs doublet 
$\varphi_u$ and $\varphi_d$ to be the complex conjugate of each other 
\be
\varphi_u= \varepsilon \cdot \varphi_d^*
\label{nonsusyhiggs}
\ee
This is the choice in the minimal standard model. 

\subsubsection{Supersymmetric Standard Model} 
In the supersymmetric models, 
the Yukawa interaction has to come from the 
$F$-type interaction. 
This implies that the superfield participating in the Yukawa 
interaction should have the same chirality. 
Therefore the choice (\ref{nonsusyhiggs}) cannot be taken, 
since the chirality changes by complex conjugation. 
Namely the Higgs superfield $H_u$ corresponding to $\varphi_u$ 
and $H_d$ corresponding to $\varphi_d$ have to be different. 
\be
H_u \not = \varepsilon \cdot H_d^*
\ee
The supersymmetric Yukawa interaction is given by 
\be
L_{Yukawa}= -{1 \over 4}\bar D D W (\Phi) + h. c. 
\ee
\ba
&&
W
\\
&
=
&
f^{ij}_u U_{i}^c H_u^T  \varepsilon  Q_j + 
f^{ij}_d D_{i}^c H_d^T  \varepsilon  Q_j + 
f^{ij}_e E_{i}^c H_d^T  \varepsilon  L_j  
\nonumber
\ea
where we denoted the negative chiral scalar superfield by capital 
letters and the charge conjugate of the positive chiral scalar 
superfield in terms of the upper suffix $c$. 

\subsection{Particle Content} 
Now we find that we need at least a pair of Higgs doublet superfield, 
we will list the minimal particle content of the supersymmetric 
standard model. 
We shall use the convention for the $U(1)$ charge $Y$ as 
\be
Q=I_3+Y
\ee

Let us note that the Higgsino (chiral 
fermions associated with the Higgs scalar) in general introduces 
the anomaly in gauge currrents. 
The simplest way out of such anomaly problem is to introduce the 
Higgsino doublet in pairs. 
Then the anomaly coming from $\tilde \varphi_u$ and  $\tilde \varphi_d$ 
always cancel each other. 
This is another reason to introduce pair of Higgs doublet superfield 
$H_u$ and $H_d$. 

\begin{tabular}{|c|c|c|c|c|c|c|}                     \hline \hline
   & $J=1$ & $J=1/2$ & $J=0$ & I & Y & $SU(3)$ 
\\ \hline \hline 
Gauge fields &  &  &  & & &   
\\ \hline 
$G$ & $g_{\mu}$ & $\tilde g$ & & & & \\
$W$ & $W_{\mu}$ & $\tilde W$ & & & & \\
$B$ & $B_{\mu}$ & $\tilde B$ & & & & \\ \hline \hline 
Higgs field & & & & & & \\\hline 
$H_u = 
\left(
\begin{array}{c}
H_u^+ \\
H_u^0 \\
\end{array}
\right)$ 
& & $\tilde \varphi_u$ & $\varphi_u$ & ${1 \over 2}$ & ${1 \over 2}$ 
&    \\ 
$H_d =
\left(
\begin{array}{c}
H_d^0 \\
H_d^- \\
\end{array}
\right)$ 
& & $\tilde \varphi_d$ & $\varphi_d$ & ${1 \over 2}$ & $-{1 \over 2}$ 
&    
\\ \hline \hline 
Quark field &  & & & & & \\ \hline
$Q_i =
\left(
\begin{array}{c}
U_i \\
D_i \\
\end{array}
\right)$ 
& & $q_i$ & $\tilde q_i$ & ${1 \over 2}$ & ${1 \over 6}$ & $3$  \\ 
$U_i^c$ & & $u_i^c$ & $\tilde u_i^c$ & 0 & $-{2 \over 3}$ & $3^*$ \\  
$D_i^c$ & & $d_i^c$ & $\tilde d_i^c$ & 0 & ${1 \over 3}$ & $3^*$  
\\ \hline \hline
Lepton field &  & & & & & \\ \hline
$L_i =
\left(
\begin{array}{c}
N_i \\
E_i \\
\end{array}
\right)$ 
& & $l_i$ & $\tilde l_i$ & ${1 \over 2}$ & $-{1 \over 2}$ &    \\
$E_i^c$ & & $e_i^c$ & $\tilde e_i^c$ & $0$ & $1$ &     \\
($N_i^c$ & & $\nu_i^c$ & $\tilde \nu_i^c$ & $0$ & $0$ &  )   
\\ \hline \hline
\end{tabular}

\end{document}